\documentclass[12pt]{article}

\newcommand{\p}{\bot}
\newcommand{\dd}{\partial}
\newcommand{\de}{\delta}
\newcommand{\m}{\mu}
\newcommand{\n}{\nu}
\newcommand{\ls}{\left(}
\newcommand{\rs}{\right)}
\newcommand{\po}{{\Pi_{\!\!\bot}}}
\newcommand{\na}{\nabla\!}
\newcommand{\al}{\alpha}

\newcommand{\pua}[2]{\left\{#1,#2\right\}}
\newcommand{\la}{\lambda}
\newcommand{\dz}{\zeta}
\newcommand{\tri}[1]{\mathop{#1}\limits^{\scriptscriptstyle 3}{}}

\newcommand{\disn}[2]{$$\displaylines{\refstepcounter{equation}%
            \label{#1}\hskip 1em minus 1em #2\hfilneg}$$}
\newcommand{\nom}{\hfil\hskip 1em minus 1em (\theequation)}
\newcommand{\no}{\hfil \hskip 1em minus 1em\phantom{(\theequation)}%
            \hfilneg\cr\hfilneg\hskip 1em minus 1em\hfil}
\newcommand{\nss}{\hfill\cr\hfill}

\makeatletter
\renewcommand{\section}{\@startsection{section}{1}{0pt}%
          {3.5ex plus 1ex minus .2ex}{2.3ex plus .2ex}{\noindent\hfil\bf}}
\makeatother

\begin{document}

\title{Constraint algebra for Regge-Teitelboim\\ formulation of gravity}

\author{S.A.~Paston\thanks{E-mail: paston@pobox.spbu.ru},
A.N.~Semenova\thanks{E-mail: cat.86@mail.ru}\\
{\it Saint Petersburg State University, Russia}
}
\date{\vskip 15mm}
\maketitle

\begin{abstract}
We consider the formulation of the gravity theory first suggested by Regge and
Teitelboim where the space-time is a four-dimensional surface in a flat
ten-dimen\-sional space. We investigate a canonical formalism for this theory following the
approach suggested by Regge and Teitelboim.
Under constructing the canonical formalism we impose additional
constraints agreed with the equations of motion.
We obtain the exact form of the first-class constraint algebra. We show that this
algebra contains four constraints which form a subalgebra (the ideal), and if these
constraints are fulfilled, the algebra becomes the constraint algebra of the
Arnowitt-Deser-Misner formalism of Einstein's gravity. The reasons for the existence of
additional first-class constraints in the canonical formalism are discussed.
\end{abstract}

\newpage

\section{Introduction}
More than 30 years ago, in 1975 T.~Regge and C.~Teitelboim
suggested  a formulation of gravity \cite{regge} similar to
the formulation of string theory. They assumed  that our space-time
is a four-dimensional surface in ten-dimensional Minkowski space
$R^{1,9}$ with one timelike and nine spacelike dimensions. In
this case the variables describing the gravity are the embedding
function of this surface into the ambient space. The authors take the standard
Einstein-Hilbert expression for the action. In this
expression they replace the metric by the induced metric
expressed in terms of the embedding function. We will call this
formulation of gravity the embedding theory. In this approach the
equations of motion (the "Regge-Teitelboim equations") appear to be
more general than the Einstein equations and they contain extra solutions.

To overcome the problem of extra solutions T.~Regge and
C.~Teitelboim suggested in \cite{regge} to impose additional
constraints $G_{\m\p}=0$ ("Einstein's constraints"), where
$G_{\m\n}$ is the Einstein tensor, $\m,\n,\ldots=0,1,2,3$, and
the symbol $\p$ denotes the direction orthogonal to the
constant time surface. While constructing the canonical formalism
these constraints are considered similarly to the primary
constrains, resulting in a system of eight constraints. We
will call the theory arising from such approach the
Regge-Teitelboim formulation of gravity, in contrast to  the embedding theory.

The approach to the gravity based on the consideration of a surface
in a flat Minkow\-ski space could be more convenient than a
standard approach when we try to develop a quantum theory of
gravity, since in this case we have  a possibility to formulate the causality
principle more clearly. In the quantum field theory the causality principle usually
means  the commutativity of operators related
to areas separated by a spacelike interval.
This principle is difficult to formulate in the framework of the standard gravity
formulation in terms of metric $g_{\m\n}$, because the interval
between points is determined by the metric which is an
operator itself. Therefore for two specific points of the space-time it is impossible to
determine what kind of interval separates them
independently of a specific  state. In the case of the description of
gravity as a dynamics of a three-dimensional surface in a flat
ambient space, we can try to work out a quantum field theory
giving this gravity in the classical limit. In this case  the problem of formulation of the
causality principle would be
solved, since the causality in the flat ambient space can be
determined by standard means of quantum field theory.

The Regge-Teitelboim formulation of gravity
has been discussed in the  work \cite{deser}
published immediately after \cite{regge}.
The authors remarked that the Regge-Teitelboim equations
are trilinear in the second derivatives of the embedding function.
This fact is very significant for this approach
as it obstructs linearizing the theory near the flat surface.
Also in \cite{deser} the problem of absence of uniqueness of the embedding is
discussed.
The lack of uniqueness causes a question whether  the transfer from one surface to
another with the same
metric corresponds to  the change of some physical degrees of freedom, or such a
transfer should be
considered as "a change of the embedding gauge". The discussion of these problems
is beyond the scope of our article.

In the paper \cite{deser} it is also noted that an artificial, {\it ad hoc},
imposing of additional constraints to the theory seems not
quite satisfactory and that a more satisfactory alternative would
be to find another action whose Euler-Lagrange equation would be
equivalent to Einstein equations. Such an action was suggested in the paper
\cite{tmf07}. In Section~4 of our paper we clarify the way of building the action and
discuss the meaning of the existence of additional first-class constraints in the
canonical formalism.

After the articles \cite{regge,deser} the idea of embedding was used
for description of gravity quite often. In particular, the
canonical formalism for the embedding theory without imposing
Einstein's constraints was investigated in \cite{tapia,frtap,rojas06}.
Such a canonical formalism turns out to be very complicated. Among
recent works using the idea of embedding we mark
\cite{faddeev1,faddeev2,rojas09}. An extended bibliography related to the
embedding theory and similar problems can be found in \cite{tapiaob}.

In the work \cite{regge} the form of the constraints system for
Regge-Teitelboim formulation of gravity was found. Also the problem
has been formulated to investigate the algebra of these constraints
and to verify whether these constraints are the first-class
constraints. However this problem is not completely solved by now.
It is probably due to the fact that one of constraints in
\cite{regge} was written incorrectly, as it was shown in \cite{tmf07},
see details in Sec.~2.

We started to  work on this problem in the article \cite{sbshk05}. It
was analyzed in detail under what conditions the imposition of
Einstein's constraints turns the Regge-Teitelboim equations into the
Einstein's equations. This is true in generic case, i.~e.,
except some special values of variables at a fixed instant. The
canonical formalism for Regge-Teitelboim formulation of gravity
was built anew in \cite{tmf07} with a supplementary imposition of Einstein's
constraints. A correct form of all constraints was obtained, however the constraint
algebra has not been
found completely. In this paper we are completing the
solution of the problem. We perform an accurate calculation
of the Poisson brackets between constraints and, as a result,
we obtain a first-class constraint algebra for Regge-Teitelboim
formulation of gravity.

For convenience we describe in Section~2  the
construction of the canonical formalism for Regge-Teitelboim
formulation of gravity following \cite{tmf07}. We do it in
particular in order to explain why we regard that in \cite{regge}
one of constraints was written incorrectly and to show how we
obtain a correct form of all constraints.

Section~3 contains the main result of this paper. In this section
we find all Poisson brackets between constraints and we obtain
the first-class constraint algebra. We also discuss the relation
between this algebra and the constraint algebra of
Arnowitt-Deser-Misner formalism for the  Einstein's gravity. The
formalism used in calculations is described in
\cite{tmf07,sbshk05}.

\section{Canonical formalism with additionally imposed Einstein's constraints}
In this section we build a canonical formalism for Regge-Teitelboim
formulation of gravity following \cite{tmf07}. We additionally
impose Einstein's constraints as it was suggested in \cite{regge}.

The embedding function determining the four-dimensional surface
$W^4$ in the flat ten-dimensional space $R^{1,9}$ is the map
 \disn{v3.2}{
y^a(x^\m):R^4\longrightarrow R^{1,9}.
\nom}
Here and below, the indices $a,b,\dots$  run over the values
$0,1,2,\dots,9$; and  $y^a$  are the Lorentzian coordinates in $R^{1,9}$.
A constant metric $\eta_{ab}=diag(1,-1,-1,\dots,-1)$
in the ambient space $R^{1,9}$   can easily rise and lower indices.
It induces on the surface $W^4$ the metric
 \disn{2}{
g_{\m\n}=\eta_{ab}\,\dd_\m y^a\, \dd_\n y^b=\eta_{ab}\,e^a_\m e^b_\n,
\nom}
where $e^a_\m\equiv\dd_\m y^a$.

We take the theory action in the form of the standard Einstein-Hilbert expression
 \disn{39d}{
S=\int d^4x\, \sqrt{-g}\;R,
\nom}
 where we substitute the induced metric expressed in terms  of the
embedding function $y^a(x)$ by formula (\ref{2}). We consider the
gravity with matter absent, because adding  matter does
not play a fundamental role in the analysis of the theory.

Note that the issue of changing  the physical content of the theory
under  a non-point change of variables including
time derivatives was studied in paper \cite{gitman}.
It  has been shown, under some assumptions, that if after the change of
variables the higher derivatives are contained in  the Lagrangian only in the form of a
combination being  a total time derivative, then the physical content of the theory
remains unchanged.
 Substituting (\ref{2}) in the action (\ref{39d}), the above condition  is fulfilled (see
below).
Nevertheless in this case the result \cite{gitman} is inapplicable, as the assumptions
made there are violated.
In particular, the change of variables (\ref{2}) is quadratic
whereas in the paper \cite{gitman} only  infinitesimal transformations are considered.

Varying action (\ref{39d}) with respect to $y^a(x)$ gives the
Regge-Teitelboim equations which can be written as
 \disn{50}{
G^{\m\n}\,b^a_{\m\n}=0,
\nom}
where $G^{\m\n}$ is Einstein's tensor, and
 \disn{22}{
b^a_{\m\n}=\po^a_b\,\dd_\m\dd_\n y^b=\na_\m e^a_\n
\nom}
is the second fundamental form of the surface. Here $\na_\m$ is
the covariant derivative, and the quantity $\po^a_b$ is the
projector on the space orthogonal to the surface $W^4$ at a given
point. We note that although  the free index $a$ ranges over 10 values,
there are only 6 independent equations, and the rest 4 equations
satisfy identically because of the properties of the second
fundamental form of the surface.

Besides the solutions of Einstein equations $G^{\m\n}=0$,
the equations (\ref{50}) contain extra solutions which can be
excluded (in general case) by imposing at the initial instant
the Einstein's constraints
 \disn{50.1}{
n_\m G^{\m\n}=0,
\nom}
where $n_\m$ is a unit vector normal to surfaces
$x^0=const$ at each point (see \cite{tmf07,sbshk05}).

For developing a canonical formalism it is convenient to drop the
total derivative term in the integrand in (\ref{39d}) and to rewrite
the action under the Arnowitt-Deser-Misner (ADM) form \cite{adm}:
 \disn{77.1}{
S=\int d^4x\, \sqrt{-g}\ls(K^i_i)^2-K_{ik} K^{ik}+\tri{R}\rs,
\nom}
where $K_{ik}$ is the second fundamental form of the surface
$t=const$ considered as a submanifold in $W^4$. Here and below,
the indices $i,k,\dots$ range the values $1,2,3$, and we label the
quantities related to the surface $t=const$ with the digit "3"
over the letter.

If we rewrite this expression in terms of the embedding function $y^a(x)$
it becomes
 \disn{77.2}{
S=\int d^4x \sqrt{-g}\ls
n_a\, n_b\;
\tri{b}^a_{ik}\,\tri{b}^b_{lm}
L^{ik,lm}+\tri{R}\rs,
\nom}
where we introduced
\disn{t7}{
L^{ik,lm}=\tri{g}^{ik}\tri{g}^{lm}-
\frac{1}{2}\ls\tri{g}^{il}\tri{g}^{km}+\tri{g}^{im}\tri{g}^{kl}\rs,\quad
L^{ik,lm}=L^{ki,lm}=L^{lm,ik}
\nom}
(this quantity is equal to a known  Wheeler-De Witt metric within a factor).
The action (\ref{77.2}) can be rewritten in the form where
the derivatives $\dot y^a\equiv \dd_0 y^a$ of the variables $y^a(x)$
with respect to the time $x^0$ are written explicitly:
\disn{t1}{
S=\int dx^0\, L(y^a,\dot y^a),\no
L=\int d^3x\;\frac{1}{2}\ls
\frac{\dot y^a\;B_{ab}\;\dot y^b}{\sqrt{\dot y^a\;\tri{\po}_{ab}\;\dot y^b}}+
\sqrt{\dot y^a\;\tri{\po}_{ab}\;\dot y^b}\;B^c_c\rs,
\nom}
where the quantity
\disn{t2}{
B^{ab}=2\sqrt{-\tri{g}}\;\tri{b}^a_{ik}\tri{b}^b_{lm}L^{ik,lm},
\nom}
as well as the projection operator $\tri{\po}_{ab}$ do not contain time derivatives.

We find the generalized momentum $\pi_a$ for the variable
$y^a$ from action (\ref{t1}):
\disn{t10}{
\pi_a=\frac{\de L}{\de \dot y^a}=
B_{ab}n^b-\frac{1}{2}n_a\ls n_c B^{cd} n_d-B^c_c\rs,
\nom}
where we use the formula
 \disn{n3.1b}{
n^a=\frac{\tri{\po}^a_b\,\dot y^b}{\sqrt{\dot y^c\,\tri{\po}_{cd}\,\dot y^d}}.
\nom}

We suppose that besides the primary constraints appearing from
this equality, four Einstein's constraints (\ref{50.1}) are also satisfied.
As shown in \cite{tmf07,sbshk05}, they can be written as
\disn{t13}{
{\cal H}^0=\frac{1}{2}\ls n_c B^{cd} n_d-B^c_c\rs\approx 0,
\nom}
\vskip -1em
\disn{t16}{
{\cal H}^i=-2\sqrt{-\tri{g}}\;\,\tri{\na}_k\ls L^{ik,lm}
\,\tri{b}^a_{lm}\, n_a\rs\approx 0.
\nom}
We note that the definitions of the constraints (\ref{t13}) differ from
these used in \cite{tmf07} by the factor  $1/2$.

If we use the constraint (\ref{t13}) in equality (\ref{t10}), then  it takes a simple form
\disn{t17}{
\pi_a=B_{ab}n^b.
\nom}
Taking account of  (\ref{t2}) and of the properties of the quantity $\tri{b}^a_{ik}$,
we immediately obtain three primary constraints
\disn{t12}{
\Phi_i=\pi_a\tri{e}^a_i\approx 0.
\nom}
One more constraint, the fourth one, has to appear. In \cite{regge}
it was obtained from the unit vector $n^b$ normalization under the form
\disn{g2}{
\ls B^{-1}\pi\rs^2-1\approx 0,
\nom}
where $B^{-1}$ meaned the inversion of matrix $B$ in seven-dimensional
subspace normal to the surface $W^3$. However, this form was incorrect,
because the matrix $B$ has rank 6 in general case and
could not be inverted in the seven-dimensional subspace mentioned above.

Indeed, the quantity $\tri{b}^a_{ik}$ can be considered as a set
of six vectors (at fixed  values of indices $i,k$ over  which it is symmetric).
On the other hand, this quantity satisfies three identities $\tri{b}^a_{ik}\tri{e}_{a,l}=0$.
Therefore, in general case there exists  a unique vector $w_a$ determined by
conditions
\disn{t3}{
w_a\tri{e}^a_l=0,\qquad w_a\tri{b}^a_{ik}=0,\qquad |w_a w^a|=1.
\nom}
The matrix $B^{ab}$ gives a zero when it acts on this vector lying in
the seven-dimensio\-nal subspace mentioned above. Hence, it can not be
inverted in this subspace. Instead of (\ref{g2}) the fourth primary
constraint has to be written as
\disn{t18}{
\Psi^4=\pi_a w^a\approx 0
\nom}
(the reason for such notation will be obvious below)
and the condition of normalization of vector $n^b$ does not lead to
new  limitations.

Using formulas (\ref{t1}),(\ref{n3.1b}),(\ref{t17}),
one easily founds that the Hamiltonian of the theory
\disn{t18.1}{
H=\int\! d^3x\, \pi_a \dot y^a - L
\nom}
vanishes. Therefore, the generalized Hamiltonian reduces to a
linear combination of constraints
(\ref{t13}),(\ref{t16}),(\ref{t12}),(\ref{t18}).

In the canonical formalism, constraints must be expressed via
generalized coordinates and momenta, i.~e., via $y^a$ and $\pi_a$
but not $\dot y^a$ in our case. Constraints (\ref{t12}) and
(\ref{t18}) satisfy this requirement (we note that vector $w_a$
determined by conditions (\ref{t3}) depends on $y^a$, but not on
$\dot y^a$), while constraints (\ref{t13}) and (\ref{t16}) do not
satisfy it. They must therefore be transformed to the necessary
form. For this, we introduce the quantity $\al^{ik}_a$
unambiguously determined by the conditions
\disn{t20}{
\al^{ik}_a=\al^{ki}_a,\quad
\al^{ik}_a\, \tri{e}^a_l=0,\quad
\al^{ik}_a w^a=0,\quad
\al^{ik}_a\, \tri{b}^a_{lm}=\frac{1}{2}\ls\de^i_l\de^k_m+\de^i_m\de^k_l\rs.
\nom}
The value $\al^{ik}_a$ can be considered as inverse to $\tri{b}^a_{lm}$, and
\disn{t20.1}{
\al^{ik}_b\, \tri{b}^a_{ik}=\tri{\po}^a_b-\frac{w^aw_b}{w^cw_c},
\nom}
where the right part contains the projector on the
six-dimensional subspace normal to the surface $W^3$ and to the vector
$w^a$.

It is clear that $\al^{ik}_a$ as well as $w_a$ depends on $y^a$
but not on $\dot y^a$. Relation (\ref{t17}) implies that
\disn{t22}{
\tri{b}^b_{ik}n_b=\frac{1}{2\sqrt{-\tri{g}}}\;
\hat L_{ik,lm}\,\al^{lm}_a\,\pi^a,
\nom}
where
\disn{t21}{
\hat L_{pr,lm}=\frac{1}{2}\ls g_{pr}g_{lm}-g_{pl}g_{rm}-g_{pm}g_{rl}\rs,\no
\hat L_{pr,lm}L^{ik,lm}=\frac{1}{2}\ls\de^i_p\,\de^k_r+\de^i_r\,\de^k_p\rs.
\nom}
Using formula (\ref{t22}), the constraints (\ref{t13}),(\ref{t16})
can be expressed in terms of $y^a$ and $\pi_a$. It is convenient
to use a linear combination $\Psi^i={\cal
H}^i+\tri{g}^{ik}\Phi_k$ instead of the constraint ${\cal H}^i$.
As a result, we have a set of eight constraints:
\disn{t27}{
\Phi_i=\pi_a\tri{e}^a_i,\qquad
\Psi^i=-\sqrt{-\tri{g}}\;\tri{\na}_k\!\ls\frac{1}{\sqrt{-
\tri{g}}}\;\pi^a\al_a^{ik}\rs+\pi^a\tri{e}_a^i,\qquad
\Psi^4=\pi_a w^a,\no
{\cal H}^0=\frac{1}{4\sqrt{-\tri{g}}}\;\pi^a\al_a^{ik}\hat L_{ik,lm}\al^{lm}_b\pi^b-\sqrt{-
\tri{g}}\;\tri{R}.
\nom}
We  can see that all constraints except ${\cal H}^0$ are linear
in momentum $\pi^a$, and the constraint ${\cal H}^0$  is quadratic.

\section{Constraint algebra}
In this section we find the exact  form of all Poisson brackets
between the constraints. It will be seen that these Poisson
brackets are linear combinations of the constraints, therefore this set of eight
constraints forms  a first-class constraint algebra
for Regge-Teitelboim formulation of gravity. We drop some
tedious algebraic transformations using formulas described in \cite{tmf07,sbshk05}.

It is convenient to work with constraints convoluted with arbitrary functions.
It is also  convenient to join constraints $\Psi^i$ and $\Psi^4$ under the index $A$
ranging  the values $1,2,3,4$, since, as it will be shown, their
action on variables has similar geometrical meaning in spite of their
different nature ($\Psi^4$ is a primary
constraint and $\Psi^i$ contains an additionally imposed constraint
${\cal H}^i$). We use denotations
 \disn{g3}{
\Phi_\xi\equiv\int d^3x\; \Phi_i(x)\,\xi^i(x)=\int d^3x\; \pi_a\tri{e}^a_i\xi^i,\qquad
{\cal H}^0_\xi\equiv\int d^3x\; {\cal H}^0(x)\,\xi(x),\no
\Psi_\xi\equiv\int d^3x\; \Psi^A(x)\,\xi_A(x)=\int d^3x\;\pi^a\ls
\al_a^{ik}\tri{\na}_i\xi_k+\tri{e}^k_a\xi_k+w_a\xi_4\rs=\nss=
\int d^3x\;\pi^a V_a^A\xi_A,
\nom}
where a denotation for the differential operator is used:
 \disn{g4}{
V_a^i=\al_a^{ik}\tri{\na}_i+\tri{e}^k_a,\qquad
V_a^4=w_a.
\nom}

First of all, we find a geometrical meaning of three constraints $\Phi_i$.
For this purpose we calculate their action on variables. It is easy to find that
\disn{t51}{
\pua{\Phi_\xi}{y^a(x)}=
\xi^i(x)\dd_i y^a(x),\qquad
\pua{\Phi_\xi}{\frac{\pi_a(x)}{\sqrt{-\tri{g}(x)}}}=
\xi^i(x)\dd_i \frac{\pi_a(x)}{\sqrt{-\tri{g}(x)}},
\nom}
where $\{\dots\}$ is a Poisson bracket. It means that $\Phi_\xi$
generates a transformation $x^i\to x^i+\xi^i(x)$ of
three-dimensional coordinates on the constant-time surface $W^3$
(it should be noted that generalized momentum $\pi^a$ is a
three-dimensional scalar density). Because all constraints
(\ref{g3}) are covariant (in three-dimensional meaning)
equalities, we can write the action of constraints $\Phi_i$ on them:
 \disn{g5}{
\pua{\Phi_\xi}{\Phi_\dz}=-\int d^3x\;\Phi_k\ls\xi^i\tri{\na}_i\dz^k-\dz^i\tri{\na}_i\xi^k\rs,
\nom}\vskip -1em
 \disn{g5.1}{
\pua{\Phi_\xi}{\Psi_\dz}=-\int
d^3x\ls\Psi^k\ls\xi^i\tri{\na}_i\dz_k+\dz_i\tri{\na}_k\xi^i\rs+\Psi^4\,\xi^i\dd_i\dz_4\rs,
\nom}\vskip -1em
 \disn{g5.2}{
\pua{\Phi_\xi}{{\cal H}^0_\dz}=-\int d^3x\;{\cal H}^0\,\xi^i\dd_i\dz.
\nom}

Now we find a geometrical meaning of four constraints $\Psi^A$.
It is easy to verify that
 \disn{g6}{
\pua{\Psi_\xi}{\tri{g}_{ik}(x)}=0,
\nom}
so the constraints $\Psi^A$ generate transformations which are an
isometric bending of the surface $W^3$ (we stress that it is true
as for $\Psi^i$ so for $\Psi^4$).  We note that the number
(four) of the found generators of three-dimensional isometric
bendings corresponds to the difference between the
dimensionality (ten) of the space into which the three-dimensional
surface is embedded and the number of independent
components (six) of the three-dimensional metric.

It is useful to calculate the action of constraints $\Psi^A$ on quantity
 \disn{g6.1}{
\pi^{lm}\equiv-\pi^a\al_a^{lm}/2.
\nom}
The calculation gives a rather long  equality where  each term is
proportional to one of constraints $\Psi^A$. Thus, under the action of
$\Psi^A$ the quantity  $\pi^{lm}$ does not change  if $\Psi^A=0$.
Since ${\cal H}^0$ and ${\cal H}^i$ can be expressed by
quantities $\tri{g}_{lm}$ and $\pi^{lm}$ (see~(\ref{t27})), we
can at once conclude (taking in to account (\ref{g5.1})) that the
Poisson bracket of constraint $\Psi_\xi$ with constraints
$\Psi^i$ and ${\cal H}^0$ reduces to a linear combination of constraints.
After  tedious calculations we
obtain an exact  result of the action of constraints $\Psi^A$ on
other constraints:
 \disn{g8}{
\pua{\Psi_\xi}{\Psi_\dz}=\int d^3x\;\ls\de y^a_{\Psi_\xi} \,\overline\Psi_{ab}\;\de
y^b_{\Psi_\dz}-
\de y^a_{\Psi_\dz} \,\overline\Psi_{ab}\;\de y^b_{\Psi_\xi}\rs,
\nom}\vskip -1em
 \disn{g9}{
\pua{\Psi_\xi}{{\cal H}^0_\dz}=\int d^3x\;\ls\de y^a_{\Psi_\xi} \,\overline\Psi_{ab}\;\de
y^b_{{\cal H}^0_\dz}-
\de y^a_{{\cal H}^0_\dz} \,\overline\Psi_{ab}\;\de y^b_{\Psi_\xi}\rs,
\nom}
where the quantity
 \disn{g10.1}{
\overline\Psi_{ab}=\ls\Psi^i\eta_{ab}-\Psi^4\frac{w_b}{w_c w^c} V^i_a\rs \tri{\na}_i
\nom}
is a linear combination of the constraints $\Psi^A$, being also  (like
$V_a^A$, see~(\ref{g4})) a differential operator. We have denoted
 \disn{g10.2}{
\de y^a_{\Psi_\xi}(x)=\pua{\Psi_\xi}{y^a(x)}=V^{aA}\xi_A(x),\quad
\de y^a_{{\cal H}^0_\dz}(x)=\pua{{\cal H}^0_\dz}{y^a(x)}=\hat B^{ac}\pi_c\dz
\nom}
for the results of acting of constraints on the independent variable $y^a(x)$, where
 \disn{g11}{
\hat B^{ac}=\frac{1}{2\sqrt{-\tri{g}}}\;\al^a_{ik}\al^c_{lm}\hat L^{ik,lm}
\nom}
is the inverted quantity to $B_{cb}$ in six-dimensional subspace
normal to the surface $W^3$ and to the vector $w^a$:
 \disn{g12}{
\hat B^{ac}B_{cb}=\tri{\po}^a_b-\frac{w^aw_b}{w^cw_c}
\nom}
(formulas (\ref{t20}),(\ref{t20.1}),(\ref{t21}) are used).

In order to complete finding of the full constraint algebra
we need to calculate the Poisson bracket of the constraint
${\cal H}^0$ with itself. This calculation is the most
tedious and gives:
 \disn{g13}{
\pua{{\cal H}^0_\xi}{{\cal H}^0_\dz}=
\int d^3x\Biggl(
\de y^a_{{\cal H}^0_\xi}\,\overline\Psi_{ab}\;\de y^b_{{\cal H}^0_\dz}\,-\,
\de y^a_{{\cal H}^0_\dz}\,\overline\Psi_{ab}\;\de y^b_{{\cal H}^0_\xi}\,-\,
\ls\Psi^k-\tri{g}^{kl}\Phi_l\rs\!\ls\xi\tri{\na}_k\dz-\dz\tri{\na}_k\xi\rs\!\Biggr).
\nom}
The formulas
(\ref{g5})-(\ref{g5.1}),(\ref{g8}),(\ref{g9}),(\ref{g13}) gives
the exact form of the first-class constraint algebra for
Regge-Teitelboim formulation of gravity. It should be noted that
the results of calculation of Poisson brackets (\ref{g8}),(\ref{g9})
and partially (\ref{g13})  have similar structure. The reason
for  that is unclear.

According to what is written after the formula (\ref{t18.1}), the
generalized Hamiltonian of the theory can be written in the form
\disn{g14}{
H^{\mbox{\scriptsize gen}}=\int\! d^3x\ls \tilde\la^i\Phi_i+N_A\Psi^A+N_0{\cal H}^0\rs.
\nom}
As can be seen from (\ref{g8}) (taking into account
(\ref{g10.1})), the four constraints $\Psi^A$ generating
the isometric bending of the surface $W^3$ form a subalgebra in the full
constraint algebra. It means that the Poisson brackets between
them are reduced to their  linear combination. Moreover,
it is seen that the Poisson brackets of the constraints $\Psi^A$ with all
other constraints  (and consequently with the Hamiltonian (\ref{g14}))
also reduce to such a linear combination. Thus the constraints
$\Psi^A$ form the ideal. It means that once imposed the constraints
$\Psi^A$ remain satisfied in the time independently of satisfying
other constraints.

If we limit our consideration of the system with satisfied constraints
$\Psi^A=0$, then its dynamics will be determined by the Hamiltonian
\disn{g15}{
\tilde H=\int\! d^3x\ls \tilde\la^i\Phi_i+N_0{\cal H}^0\rs=
\int\! d^3x\ls -\tilde\la_i{\cal H}^i+N_0{\cal H}^0\rs=\nss=
\int\! d^3x\ls\! -2\tilde\la_i
\sqrt{-\tri{g}}\;\tri{\na}_k\!\ls
\frac{\pi^{ik}}{\sqrt{-\tri{g}}}\rs\!+N_0\!\ls
\frac{\pi^{ik}\hat L_{ik,lm}\pi^{lm}}{\sqrt{-\tri{g}}}-
\sqrt{-\tri{g}}\;\tri{R}\rs\!\!\rs,
\nom}
where $\Phi_i$ was expressed by ${\cal H}^i$,  the formulas
(\ref{t27}) were applied, and the quantity
$\pi^{ik}$ determined by (\ref{g6.1})  was used.
This Hamiltonian as a functional of the quantities $\tri{g}_{ik}$ and $\pi^{ik}$
coincides exactly with the known Hamiltonian expression in the ADM formalism.
Besides, it is easy to verify that these quantities $\tri{g}_{ik}$ and $\pi^{ik}$
are canonically conjugate  at $\Psi^A=0$ (it should be
noted that this condition is necessary only for vanishing of the Poisson
bracket $\pua{\pi^{ik}(x)}{\pi^{lm}(\tilde x)}$ ). Therefore
the dynamics of Regge-Teitelboim formulation of gravity on the surface of constraints
$\Psi^A=0$ coincides with the dynamics of gravity in the ADM formalism.

\section{Discussion about existence of additional first-class constraints}
In this section we discuss what could  mean
the existence in the canonical formalism of additional constraints
which are in involution with the Hamiltonian of the theory
and which form a first-class constraint algebra, probably with
other constraints inherent in the theory. The
Einstein's constraints (\ref{t13}),(\ref{t16}) for Regge-Teitelboim formulation of
gravity are just these additional constraints.

For comparison we consider a simple model in the Minkowski space with the action
\disn{g16}{
S=\int\!dt\!\int\! d^3x\ls
\frac{1}{2}(\dd_0A_i)(\dd_0A_i)-\frac{1}{4}\ls\dd_iA_k-\dd_kA_i\rs\ls\dd_iA_k-
\dd_kA_i\rs\rs,
\nom}
where the  independent variable is a three-component field $A_i(x)$.
The generalized momentum is the quantity $\pi_i=\dd_0A_i$,
the primary constraints are absent. The Hamiltonian has form
\disn{g17}{
H=\int\! d^3x\ls
\frac{1}{2}\pi_i\pi_i+\frac{1}{4}\ls\dd_iA_k-\dd_kA_i\rs\ls\dd_iA_k-\dd_kA_i\rs\rs.
\nom}
We consider an additional constraint $\Phi(x)=\dd_i\pi_i(x)$. It
is easy to verify that it is in involution with the Hamiltonian, so
their Poisson bracket $\pua{H}{\Phi(x)}=0$ vanishes.
Because of $\pua{\Phi(x)}{\Phi(y)}=0$, the quantity $\Phi(x)$ is the first-class
constraint and can be added to the Hamiltonian with a Lagrange factor:
\disn{g18}{
H^{\mbox{\scriptsize gen}}=\int\! d^3x\ls
\frac{1}{2}\pi_i\pi_i+\frac{1}{4}\ls\dd_iA_k-\dd_kA_i\rs\ls\dd_iA_k-\dd_kA_i\rs
+\la\,\dd_i\pi_i\rs.
\nom}
Therefore, the case of the additional imposed constraint
$\Phi(x)$ in this model is completely analogous to the case of
Einstein's constraints in Regge-Teitelboim formulation of gravity.

We construct the action $S'$ corresponding to the Hamiltonian
(\ref{g18}). A new equality for the generalized velocity reads
\disn{g19}{
\dd_0 A_i=\frac{\de H^{\mbox{\scriptsize gen}}}{\dd\pi_i}=\pi_i-\dd_i\la.
\nom}
Expressing the generalized momentum $\pi_i$ from this equality
and making the Legendre transform we find the required action
\disn{g20}{
S'=\int\!dt\!\int\! d^3x\biggl(
\frac{1}{2}(\dd_0A_i+\dd_i\la)(\dd_0A_i+\dd_i\la)-\frac{1}{4}\ls\dd_iA_k-
\dd_kA_i\rs\ls\dd_iA_k-\dd_kA_i\rs\biggl).
\nom}
The Lagrange factor $\la(x)$ related with the additionally imposed
constraint appears in this action as a new independent variable.
Denoting  $A_0=-\la$ one  easy recognizes in
expression (\ref{g20}) the free electrodynamics action. The
initial action (\ref{g16}) can be derived from it by fixing of
the gauge $A_0=0$.

This example shows that the existence of additional first-class
constraints in the canonical formalism can indicate
that the initial theory without additional constraints is a
result of fixing of gauge (probably partial) in some extended
theory with an  additional gauge symmetry. In particular, the
initial embedding theory with action (\ref{39d}) having a
four-parameter gauge group, where the independent variable is the
embedding function, appears to be the result of the gauge fixing
in Regge-Teitelboim formulation of gravity which has an eight-parameter gauge group
and is described  by the Hamiltonian (\ref{g14}).
It should be noted that, as well known, the fixing of gauge in action usually leads to
loss of some equations of motion. That is why the Regge-Teitelboim equations
(\ref{50}) have extra solutions.

The action of the extended theory corresponding to generalized
Hamiltonian (\ref{g14}) of Regge-Teitelboim formulation of gravity was found in
\cite{tmf07} in a way completely analogous to the one described in this section.
It can be written in the form of the initial Einstein-Hilbert action
\disn{s2}{
S=\int d^4x\, \sqrt{-g'}\;R(g'),
\nom}
if we substitute for the metric $g'_{\m\n}$ the modification of expression (\ref{2}):
\disn{s3}{
g'_{ik}=\tri{g}_{ik}=\dd_iy^a\dd_ky_a,\quad
g'_{0k}=\dd_0y^a\dd_ky_a-N_k,\quad
g'_{00}=N_0^2+g'_{0i}\,\tri{g}^{ik}\,g'_{0k}
\nom}
(whence we obtain $g'^{00}=\frac{1}{N_0^2}$; we note that these
formulas differ from formulas in \cite{tmf07} by numeric factors
because of changing of definition of constraint (\ref{t13})).
Here $N_k$ and $N_0$ are new independent variables (in addition
to $y^a$), which are transformed into Lagrange multipliers in
canonical formalism. The action (\ref{s2}) has an eight-parameter
gauge symmetry, and the value $g'_{\m\n}$ appears  to be invariant
under four of these eight transformations, which have constraints
$\Psi^A$ as generators in the canonical formalism.

If we introduce a  partial fixing of gauge by conditions
\disn{s4}{
N_0=\sqrt{\dd_0 y^a \tri{\po}_{ab}\dd_0 y^b},\qquad N_k=0,
\nom}
then the quantity $g'_{\m\n}$ coincides with the induced metric,
and the action of the extended theory (\ref{s2}) transforms into
the action (\ref{39d}) of the initial embedding theory. If we do
not fix the gauge, then the quantity $g'_{\m\n}$ still satisfies
(in general case) the Einstein's equations. Therefore we can consider the
quantity $g'_{\m\n}$ to be the metric, which is invariant under
additional symmetry transformation and coincides with the induced
metric only in the mentioned gauge.

A disadvantage of the action (\ref{s2}) of the extended theory is the
presence of a singled out direction of time related to the fact that the time and space
components appear in formulas (\ref{s3}) in a different way.
It would be interesting to find a modification of formulas (\ref{s3})
without singled out time direction but where the equations of motion
still would be equivalent to the Einstein's equations.

\vskip 0.2em
{\bf Acknowledgments.}
The work was supported by the Russian Ministry of Education, Grant
No.~RNP.2.1.1/1575.

\end{document}